\RequirePackage{fix-cm}
\documentclass{svjour3}
\smartqed
\usepackage[paperwidth=210mm,paperheight=279mm,textwidth=156mm,left=27mm,top=32mm,textheight=226mm,headheight=12pt,headsep=9mm]{geometry}

\usepackage{graphicx}
\usepackage{amsmath}
\usepackage[T1]{fontenc}
\usepackage{newtxtext}
\usepackage[slantedGreek]{newtxmath}
\usepackage{mathrsfs}
\usepackage{booktabs}
\usepackage{url}
\usepackage{lastpage}
\usepackage{marvosym}
\usepackage[numbers,sort&compress]{natbib}
\usepackage[colorlinks=true,linkcolor=blue,citecolor=blue,urlcolor=blue]{hyperref}

\journalname{Few-Body Systems}

\makeatletter

\def\runheadhook{\rlap{\smash{\lower11pt\hbox to\textwidth{\hrulefill}}}}
\def\ps@headings{\let\@oddfoot\@empty\let\@evenfoot\@empty
  \def\@evenhead{\small\csname runheadhook\endcsname
    Page \thepage\ of \pageref*{LastPage}\hfil\leftmark\unskip}%
  \def\@oddhead{\small\csname runheadhook\endcsname
    \ignorespaces\rightmark\hfil Page \thepage\ of \pageref*{LastPage}}%
  \let\@mkboth\@gobbletwo\let\sectionmark\@gobble\let\subsectionmark\@gobble}
\def\@maketitle{\newpage
  \normalfont
  \vbox to0pt{\vskip-\headsep\vskip-15pt\nointerlineskip
    \hbox to\textwidth{\vbox{\small\baselineskip11pt
      \hbox{\@journalname\ manuscript No.}\hbox{(will be inserted by the editor)}}\hfil}%
    \vskip5pt\hrule\@height0.4pt\@width\textwidth\vss}\nointerlineskip
  \gdef\footnoterule{\kern-3\p@\hrule\@width\textwidth\kern2.6\p@}%
  {\setbox0=\vbox{\setcounter{auth}{1}\def\and{\stepcounter{auth} }%
     \hfuzz=2\textwidth\def\thanks##1{}\@author}%
   \global\value{inst}=\value{auth}}%
  \setcounter{auth}{1}%
  \vskip18mm
  {\fontsize{10.5}{13}\selectfont\bfseries\raggedright
   \def\and{\unskip\nobreak\enskip{\boldmath$\cdot$}\enskip\ignorespaces}%
   \noindent\ignorespaces\@author\par}%
  \vskip8pt
  {\fontsize{17}{20}\selectfont\bfseries\raggedright\noindent\ignorespaces\@title\par}%
  \vskip27mm
  \noindent{\small\@date\par}\vskip12mm
  \global\@minipagetrue
  \global\everypar{\global\@minipagefalse\global\everypar{}}}
\makeatother

\newcommand{\fpanel}[4]{%
  \begin{minipage}[t]{0.485\textwidth}\centering
  \makebox[\linewidth]{\makebox[0pt][l]{\textsf{\large #1}}\hfill\textsf{#2}\hfill}\\[0.5ex]
  #3\\[-0.1ex]
  \textsf{\small #4}%
  \end{minipage}}

\begin{document}

\title{Nucleon light-front transverse densities from three-body and quark\,+\,diquark Faddeev equations}
\titlerunning{Nucleon transverse densities: three-body and quark\,+\,diquark pictures}

\author{Z.-N. Xu \and Z.-Q. Yao \and P. Cheng}
\authorrunning{Z.-N. Xu, Z.-Q. Yao, P. Cheng}

\institute{Z.-N. Xu (\Letter) \at
  Dpto.\ Ciencias Integradas, Centro de Estudios Avanzados en F\'{\i}s., Mat.\ y Comp., Fac.\ Ciencias Experimentales, Universidad de Huelva, E-21071 Huelva, Spain \\
  \email{zhenni.xu@dci.uhu.es}
  \and
  Z.-Q. Yao (\Letter) \at
  Helmholtz-Zentrum Dresden-Rossendorf, Bautzner Landstra{\ss}e 400, D-01328 Dresden, Germany \\
  \email{z.yao@hzdr.de}
  \and
  P. Cheng (\Letter) \at
  Department of Physics, Anhui Normal University, Wuhu, Anhui 241000, China \\
  \email{pcheng@mail.ahnu.edu.cn}}

\date{Received: date / Accepted: date}

\maketitle

\begin{abstract}
Nucleon elastic electromagnetic form factors obtained from two complementary continuum Schwinger function approaches -- a Poincar\'e-covariant three-body Faddeev equation formulated in rainbow-ladder truncation and a quark\,+\,fully-interacting-diquark, $q(qq)$, Faddeev equation -- are used to chart light-front transverse charge and magnetisation densities of the proton, the neutron, and their dressed valence quarks.  Going beyond a comparison of observables, the bound-state amplitudes that underlie them are also compared.  Both amplitudes are dominated by $s$-wave structures; and in the three-body solution, the mixed-antisymmetric (mixed-symmetric) flavour component is dominated by structures with the Dirac character of scalar (axialvector) diquark correlations, so that the association between these flavour components and diquark correlations is realised explicitly.  The $q(qq)$ amplitude falls more slowly with the quark-diquark relative momentum than the three-body amplitude does with the corresponding Jacobi momentum.  The two descriptions deliver mutually compatible densities, which match expectations based on modern parametrisations of data.  Transverse-plane valence $u$- and $d$-quark Dirac radii are practically indistinguishable, whereas the $d$-quark Pauli radius is roughly 10\% larger; and the valence $d$ quark is magnetically far more active than the valence $u$ quark, a feature tied to the mixed-symmetric (axialvector-diquark) components of the nucleon amplitude.  For a $+\hat x$ polarised nucleon, positive charge is displaced in the $+\hat y$ direction, with the opposite effect for negative charge.
\end{abstract}

\section{Introduction}
\label{S:Intro}
Drawing charts of nucleon structure has long been a principal goal of hadron physics.  The Poincar\'e-invariant elastic electromagnetic form factors contain much objective information in this connection; and a great deal of electron\,+\,nucleon scattering data has been collected and analysed since the pioneering experiments described in Ref.\,\cite{Hofstadter:1956qs} -- see, \emph{e.g}., Refs.\,\cite{Perdrisat:2006hj, Punjabi:2015bba, Gao:2021sml, Cui:2022fyr}.  The data now extend to large values of spacelike squared momentum transfer, $0<Q^2 \lesssim 15\,$GeV$^2$ \cite{Gilfoyle:2018xsa, Wojtsekhowski:2020tlo}; so a coherent description requires a Poincar\'e-covariant treatment of a bound state whose constituents are highly relativistic.  Lattice-regularised QCD has made material progress in this area \cite{Alexandrou:2018sjm, Alexandrou:2023qor, Syritsyn:2025fiu}, but reaching the large-$Q^2$ domain now accessible to experiment remains difficult \cite{Syritsyn:2025fiu}.

Continuum Schwinger function methods (CSMs) \cite{Roberts:1994dr,Eichmann:2016yit, Binosi:2022djx, Ding:2022ows, Achenbach:2025kfx,Xu:2022kng} provide a viable alternative.  Regarding nucleon elastic electromagnetic form factors, two complementary CSM paths are currently in use: a direct approach based on a three-body Faddeev equation for the nucleon \cite{Eichmann:2011vu, Yao:2024uej}; and a quark\,+\,fully-interacting diquark, $q(qq)$, picture of nucleon structure \cite{Barabanov:2020jvn, Cheng:2025yij,Cheng:2026flu}.  Both are Poincar\'e-covariant and deliver form factors that cover and reach beyond the range of existing experiments.

\begin{figure}[t]
\centering
\includegraphics[width=0.42\textwidth]{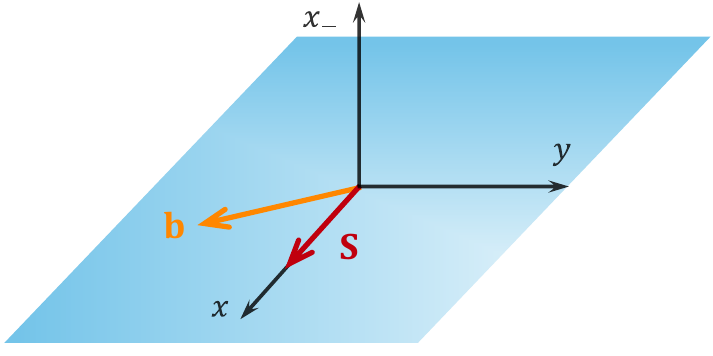}
\caption{With the nucleon's direction of motion defining the longitudinal light-front vector, the light-front transverse plane (blue surface) is characterised by the Cartesian vectors $(\hat x,\hat y)$.  $\vec{b}$ (orange) -- transverse position vector; $\vec{S}$ (red) -- transverse spin vector.  Adapted from Ref.\,\cite{Xu:2025ups}.}
\label{IMF}
\end{figure}

One particular use of such form factors is to produce charts of the distribution of charge and magnetisation within the nucleon.  Three-dimensional Fourier transforms of the Sachs form factors \cite{Sachs:1962zzc}, interpreted in the Breit frame, were long imagined to describe configuration-space densities; but Poincar\'e-covariant wave functions do not have a probability interpretation, so this identification is flawed.  A contemporary alternative \cite{Miller:2010nz} considers two-dimensional Fourier transforms of the form factors with $Q^2$ interpreted in a Drell-Yan-West frame \cite{Drell:1969km, West:1970av}.  Such functions describe true densities in the plane perpendicular to the light-front vector used in defining the nucleon state, Fig.\,\ref{IMF}.

This contribution summarises and extends Ref.\,\cite{Xu:2025ups}, wherein transverse densities were calculated and compared using both the three-body \cite{Yao:2024uej,Yao:2026phv} and $q(qq)$ \cite{Cheng:2025yij} form factors.  Whereas Ref.\,\cite{Xu:2025ups} compared observables, herein we also compare the bound-state amplitudes that underlie them, thereby exposing the structural features the two pictures share and those in which they differ.  Section~\ref{S:Amplitudes} describes the three-body and $q(qq)$ Faddeev equations and their solutions; Sec.\,\ref{S:Current} sketches the associated electromagnetic currents; Sec.\,\ref{S:density} defines the transverse densities and reports the CSM predictions; and Sec.\,\ref{S:Summary} provides a summary and perspective.

\section{Nucleon Faddeev amplitudes}
\label{S:Amplitudes}
\begin{figure}[t]
\centering
\makebox[\textwidth]{\makebox[0pt][l]{\raisebox{21mm}{\textsf{\large A}}}\hfill\includegraphics[width=0.52\textwidth]{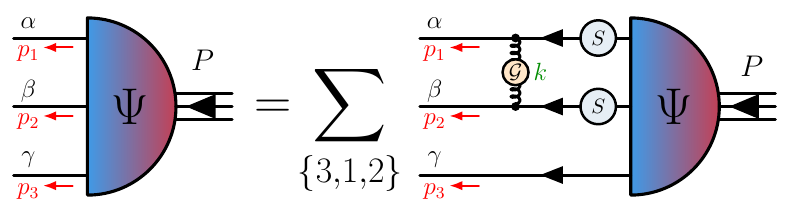}\hfill}\\[1.5ex]
\makebox[\textwidth]{\makebox[0pt][l]{\raisebox{21mm}{\textsf{\large B}}}\hfill\includegraphics[width=0.6\textwidth]{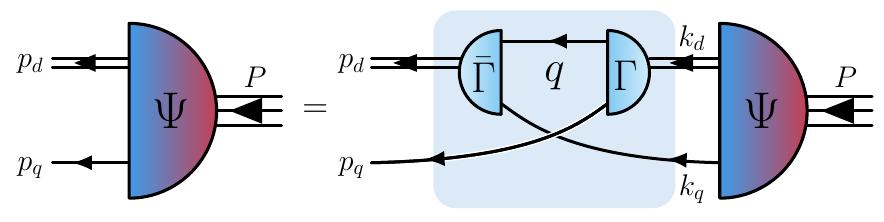}\hfill}
\caption{Faddeev equations for the nucleon.  Panel~\textsf{A}: three-body equation in RL truncation.  Half-disc: Faddeev amplitude, $\Psi$, which involves 128 independent scalar functions.  Spring: dressed-gluon interaction mediating quark\,+\,quark scattering, Eqs.\,\eqref{EqRLInteraction}, \eqref{defcalG}.  Solid line: dressed-quark propagator, $S$, calculated from the rainbow gap equation.  Lines not adorned with a shaded circle are amputated.  The sum runs over each case with quark $i=1,2,3$ as the spectator.  Isospin symmetry is assumed.  Adapted from Ref.\,\cite{Xu:2025ups}.  Panel~\textsf{B}: quark\,+\,diquark equation.  Half-disc: Faddeev amplitude, Eq.\,\eqref{qdqamp}.  Single line: dressed-quark propagator.  Double line: diquark propagator.  $\Gamma$, $\bar\Gamma$: diquark correlation amplitudes.  Shaded box: Faddeev kernel, in which binding arises from exchange of a dressed quark between the spectator quark and the diquark.}
\label{FigFaddeev}
\end{figure}

Figure~\ref{FigFaddeev} sketches the two bound-state equations considered herein: the three-body Faddeev equation (Panel~\textsf{A}), discussed in Sec.\,\ref{S:3body}; and its quark\,+\,diquark counterpart (Panel~\textsf{B}), discussed in Sec.\,\ref{S:qdq}.

\subsection{Three-body Faddeev equation}
\label{S:3body}
We work at leading order in a systematically improvable, symmetry-preserving truncation of all Dyson-Schwinger equations that appear in the matrix element which defines the photon\,+\,nucleon interaction, \emph{viz}.\ the rainbow-ladder (RL) truncation \cite{Munczek:1994zz, Bender:1996bb}.  RL truncation is known to be reliable for pion, kaon, and nucleon observables, both practically, owing to numerous successful applications \cite{Roberts:2021nhw, Ding:2022ows, Raya:2024ejx}, and because improvement schemes exist and have been tested, showing that the cumulative effect of corrections in these channels can be absorbed into a modest modification of the quark\,+\,quark scattering kernel \cite{Chang:2009zb, Binosi:2014aea, Williams:2015cvx, Binosi:2016rxz, Qin:2020jig,Xu:2022kng,Xu:2025cyj,Xu:2026sjy}.

In RL truncation, the quark\,+\,quark scattering kernel is \cite{Maris:1997tm}
\begin{equation}
\label{EqRLInteraction}
\mathscr{G}_{CD}^{AB}(k) = \tilde{\mathcal G}(k^2)\,
 \big[i\gamma_\mu\tfrac{\lambda^{a}}{2}\big]_{CB}\, \big[i\gamma_\nu\tfrac{\lambda^{a}}{2}\big]_{DA}\, T_{\mu\nu}^k\,,
\end{equation}
where $k^2 T_{\mu\nu}^k = k^2 \delta_{\mu\nu} - k_\mu k_\nu$ (Landau gauge \cite{Bashir:2009fv}) and $A,B,C,D$ denote colour and spinor indices.  We use the interaction of Refs.\,\cite{Qin:2011dd, Binosi:2014aea}:
\begin{equation}
\label{defcalG}
\tilde{\mathcal G}(y) = \frac{8\pi^2}{\omega^4} D\, {\rm e}^{-y/\omega^2} + \frac{8\pi^2 \gamma_m \mathcal{F}(y)}{\ln\big[ \tau+(1+y/\Lambda_{\rm QCD}^2)^2 \big]}\,,
\end{equation}
with $\gamma_m=12/25$, $\Lambda_{\rm QCD} = 0.234\,$GeV, $\tau={\rm e}^2-1$, ${\cal F}(y) = \{1 - \exp(-y/\Lambda_{\mathcal I}^2)\}/y$, $\Lambda_{\mathcal I}=1\,$GeV.  Asymptotic freedom is manifest in the ultraviolet behaviour of $\tilde{\mathcal G}(y)$; and confinement is expressed via the violation of reflection positivity by coloured Schwinger functions \cite{Ding:2022ows}.  A mass-independent momentum-subtraction renormalisation scheme is used \cite{Chang:2008ec}, with renormalisation scale $\zeta = 19\,$GeV.  With $\omega = 0.8\,$GeV \cite{Xu:2022kng}, $\omega D = 0.8\,{\rm GeV}^3$, and renormalisation-point-invariant current mass $\hat m_u = \hat m_d = 6.04\,$MeV (one-loop evolved mass at $\zeta=2\,$GeV of $4.19\,$MeV), one obtains $m_\pi = 0.14\,$GeV, $m_N=0.94\,$GeV, and $f_\pi=0.094\,$GeV, in good agreement with experiment \cite{ParticleDataGroup:2024cfk}.  Physical observables are practically unchanged under $\omega \to (1\pm 0.2)\omega$ with $\omega D$ held fixed \cite{Qin:2020rad}.  All results below are therefore parameter-free predictions.  The interaction in Eq.\,\eqref{defcalG} can be connected with QCD's process-independent effective charge \cite{Qin:2011dd, Cui:2019dwv, Deur:2023dzc}: it yields $\alpha_{\mathcal G}(0)/\pi = 1.45$ and a gluon mass-scale $m_{\mathcal G} = 0.54\,$GeV, which compare fairly well with the QCD values $\hat\alpha(0)/\pi = 0.97(4)$, $\hat m_0 = 0.43(1)\,$GeV \cite{Cui:2019dwv}.

The dressed-quark propagator, $S(k) = Z(k^2)/[i \gamma\cdot k + M(k^2)]$, is obtained from the RL gap equation \cite{Maris:1997tm, Maris:2005tt, Krassnigg:2008bob}.  The nucleon Faddeev amplitude, $\Psi$, then follows from the equation sketched in Fig.\,\ref{FigFaddeev}\,\textsf{A}:
\begin{align}
\Psi_{ABCD}(p,q,P) & = \sum_{i=1}^3\Psi^{(i)}_{ABCD}(p,q,P)\,, \label{faveq}\\
\Psi^{(3)}_{ABCD}(p,q,P) & = \int_{dk}\mathscr{G}^{B'A'}_{AB}(k)\, S_{A'A''}(k_1)\, S_{B'B''}(\tilde{k}_2)\, \Psi_{A''B''CD}(p^{(3)},q^{(3)},P)\,,
\label{faveq3c}
\end{align}
with $\Psi^{(1,2)}$ obtained by cyclic permutation of the quark labels.  Here $\int_{dk}$ denotes a translationally invariant regularisation of the four-dimensional integral; the indices $A,B,C,D$ now include spinor ($\alpha,\beta,\gamma,\delta$), colour ($r,s,t,u$), and flavour ($a,b,c,d$) labels, with $D$ referring to the nucleon; and $k$ is the exchanged momentum, $k_i = p_i-k$, $\tilde k_i=p_i+k$.  The quark momenta satisfy $p_1+p_2+p_3=P$, the total nucleon momentum, and we use the relative momenta \cite{Eichmann:2011vu}
\begin{equation}
p=(1-\xi)\, p_{3}-\xi\,(p_{1}+p_{2})\,, \qquad q=\tfrac{1}{2}(p_{2}-p_{1})\,,
\label{Involvesxi}
\end{equation}
with $\xi=1/3$, so that $p^{(3)}=p$ and $q^{(3)}=q+k$ in Eq.\,\eqref{faveq3c}.

Owing to its spin, flavour, and colour structure, the Faddeev amplitude can be written
\begin{equation}
\label{amp}
\Psi_{ABCD}(p,q,P) = \frac{\epsilon_{rst}}{\sqrt6}\, \sum_{\rho=0}^1 \psi_{\alpha\beta\gamma}^{\rho \, {\cal I}}(p,q,P)\, {F}^{\rho}_{abcd}\,,
\end{equation}
where the Levi-Civita symbol expresses colour antisymmetry and ${\cal I}$ is the nucleon spin label.  The flavour amplitude is expressed in a basis spanned by mixed-antisymmetric (MA) and mixed-symmetric (MS) isospin tensors:
\begin{equation}
\label{FlavourM}
\mbox{MA:}\;\; F^{0}_{abcd} = \tfrac{1}{\sqrt{2}}\, [i\sigma_2]_{ab}\, \delta_{cd}\,, \qquad
\mbox{MS:}\;\; F^{1}_{abcd} = -\tfrac{1}{\sqrt{6}}\, [\sigma_j i\sigma_2 ]_{ab}\, [\sigma_j ]_{cd}\,,
\end{equation}
where $\{\sigma_j\}$ are Pauli matrices.  Under $a\leftrightarrow b$, the MA term is antisymmetric and MS is symmetric; so, interpreted in terms of diquark-like correlations \cite{Barabanov:2020jvn}, MA may be associated with the isoscalar-scalar channel and MS with the isovector-axialvector channel.  Using an orthonormal set of $64$ Dirac-matrix-valued tensors $\{{\mathsf X}^i_{\alpha\beta\gamma{\cal I}}(p,q,P)\}$ \cite{Eichmann:2009en}, the momentum-spin components are expanded as
\begin{equation}
\label{basise}
\psi^{\rho\, {\cal I}}_{\alpha \beta \gamma}(p,q,P) = \sum_{i=1}^{64} f^{i\rho}(p^{2},q^{2},z_{0},z_{1},z_{2})\, {\mathsf X}_{\alpha\beta\gamma{\cal I}}^i(p,q,P)\,,
\end{equation}
with $z_{0}=\hat{p}_{T} \cdot \hat{q}_{T}$, $z_{1}=\hat{p} \cdot \hat{P}$, $z_{2}=\hat{q} \cdot \hat{P}$, where $\hat p^2=1$, \emph{etc}., and $p^T_\mu = T^P_{\mu\nu}p_\nu$.  Completing the colour and flavour algebra in Eq.\,\eqref{faveq3c}, one finds
\begin{align}
\psi^{\rho\,{\mathcal I}\,(3)}_{\alpha\beta\gamma}(p,q,P)
= \frac{2}{3} \int_{dk} & \tilde{\mathcal G}(k^2)\, T^k_{\mu\nu}\,
[\gamma_\mu S(k_1)]_{\alpha\alpha''}\, [\gamma_\nu S(\tilde k_2)]_{\beta\beta''} \nonumber \\
& \times \psi^{\rho\,\mathcal I}_{\alpha''\beta''\gamma}(p^{(3)},q^{(3)},P)\,.
\label{eq:psi3-spinmom}
\end{align}
Projecting with the basis in Eq.\,\eqref{basise}, one arrives at a coupled system for the $64\times 2$ scalar functions $\{f^{i\rho}\}$.  The equations for the $f^{i0}$ are decoupled from those for $f^{i1}$ in Eq.\,\eqref{eq:psi3-spinmom}, but the coupling re-emerges when permutation symmetry is used to reconstruct $\psi^{\rho\,(1,2)}$.

In interpreting the solution, it is useful to recall the structure of the basis \cite{Eichmann:2009en}.  Its elements are built from products $(\Theta_X \Gamma_i \Lambda_\sigma \gamma_5 C)_{\alpha\beta}\,(\Theta_X \Gamma_j \Lambda_+)_{\gamma{\cal I}}$, where: $X=\mathcal{S},\mathcal{P},\mathcal{V},\mathcal{A}$ labels $\Theta_X = \tfrac{1}{2}\mathbb{I}, \tfrac{1}{2}\gamma_5, \gamma^T_\mu, \gamma^T_\mu\gamma_5$, with the Lorentz index contracted between the two factors in the latter two cases; $\Lambda_\pm$ are positive- and negative-energy projection operators, $\sigma=\pm$; and $\Gamma_{i,j} \in \{\mathbb{I}, \gamma\cdot r\,\gamma\cdot s, \gamma\cdot r, \gamma\cdot s\}$, with $r$, $s$ orthonormal unit vectors spanning the components of $p$, $q$ transverse to $P$.  We denote by $\mathcal{S}^\sigma$, $\mathcal{P}^\sigma$, $\mathcal{V}^\sigma$, $\mathcal{A}^\sigma$ the elements with $\Gamma_i=\Gamma_j=\mathbb{I}$; they carry no dependence on the directions of the relative momenta, \emph{i.e}.\ they are $s$-wave structures.  Since $\gamma_5\Lambda_\sigma\gamma_5=\Lambda_{-\sigma}$, the $(12)$-pair Dirac structure of $\mathcal{S}^\sigma$ is $\Lambda_\sigma\gamma_5 C$, that of a scalar diquark, whereas $\mathcal{A}^\sigma$ involves $\gamma^T_\mu\Lambda_{-\sigma}C$, characteristic of an axialvector diquark; analogously, $\mathcal{V}^\sigma$ and $\mathcal{P}^\sigma$ involve vector-like ($\gamma^T_\mu\Lambda_\sigma\gamma_5 C$) and pseudoscalar-like ($\Lambda_{-\sigma}C$) structures.  The dependence on the angular variables is expressed via Chebyshev expansions:
\begin{equation}
\label{Cheb}
f^{i\rho}(p^2,q^2,z_0,z_1,z_2) = \sum_{n_1,n_2,n_0} c^{i\rho}_{n_1n_2n_0}(p^2,q^2)\, \tilde T_{n_1}(z_1)\, \tilde T_{n_2}(z_2)\, \tilde T_{n_0}(z_0)\,,
\end{equation}
where $\tilde T_n(z) = i^n T_n(z)$ for $n\geq 1$, $\tilde T_0 = 1/\sqrt{2}$, and $T_n$ are Chebyshev polynomials of the first kind.  In our solution, $n_{1,2}\leq 11$, $n_0\leq 7$, and the $p^2$, $q^2$ dependence is resolved on $60$ nodes extending to $40\,$GeV$^2$.

\begin{figure}[t]
\centering
\includegraphics[width=0.48\textwidth]{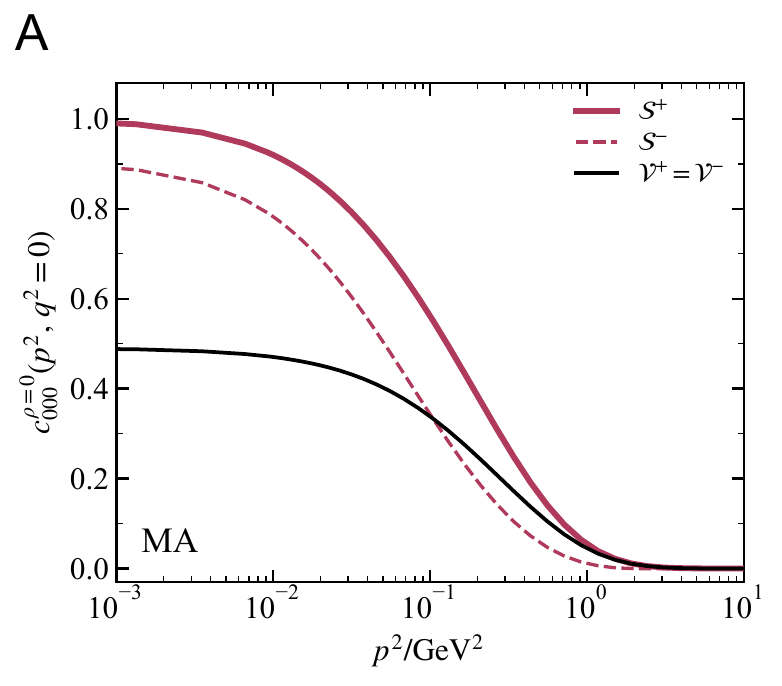}\hfill
\includegraphics[width=0.48\textwidth]{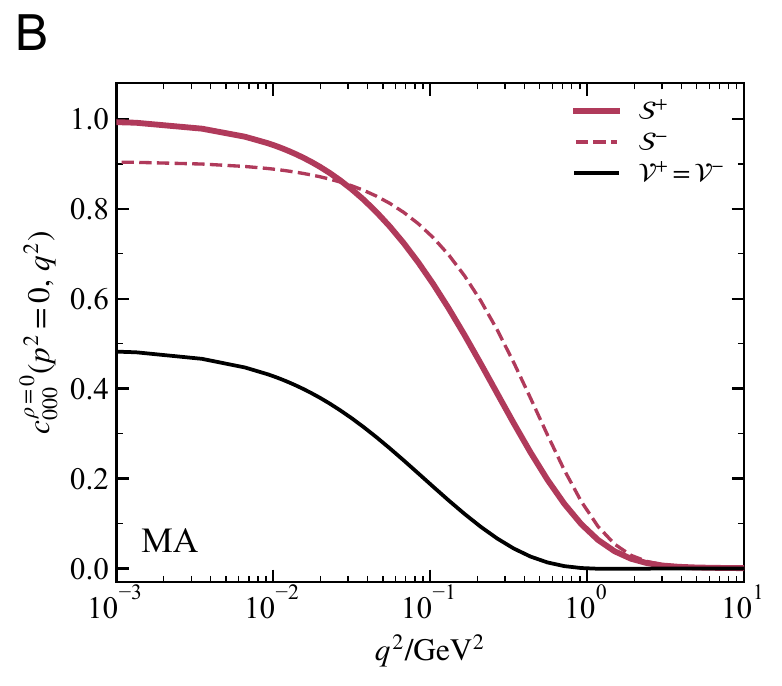}\\[1.5ex]
\includegraphics[width=0.48\textwidth]{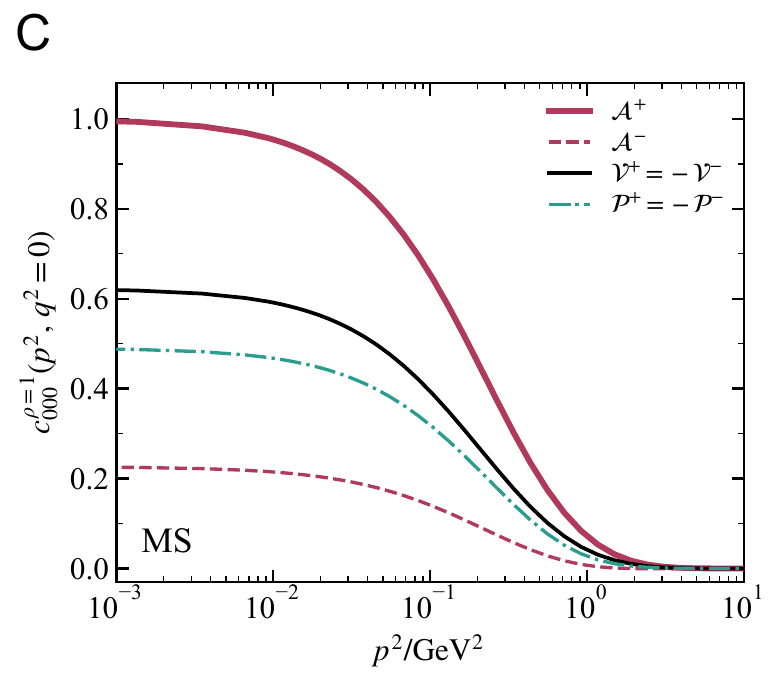}\hfill
\includegraphics[width=0.48\textwidth]{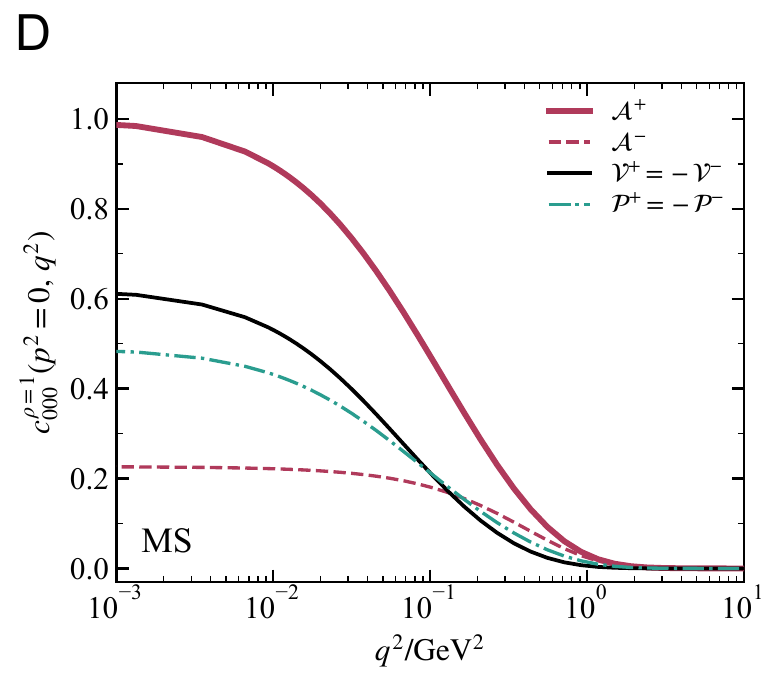}
\caption{Solution of the three-body Faddeev equation, Fig.\,\ref{FigFaddeev}\,\textsf{A}: zeroth Chebyshev moments, $c_{000}$ in Eq.\,\eqref{Cheb}, of the dominant basis components in the MA ($\rho=0$; Panels~\textsf{A}, \textsf{B}) and MS ($\rho=1$; Panels~\textsf{C}, \textsf{D}) flavour channels, as functions of $p^2$ at $q^2=0$ (\textsf{A}, \textsf{C}) and of $q^2$ at $p^2=0$ (\textsf{B}, \textsf{D}).  Thick crimson curves: leading structure, $\mathcal{S}^+$ (MA) or $\mathcal{A}^+$ (MS); dashed crimson: its $\sigma=-$ partner; black: $\mathcal{V}$; dot-dashed teal: $\mathcal{P}$.  Components related identically to those drawn are indicated in the legends.  All curves are normalised by $c_{000}(\mathcal{S}^+)$ at $p^2=q^2=0$.}
\label{fig:amp3b}
\end{figure}

The solution is illustrated in Fig.\,\ref{fig:amp3b}.  In both flavour channels, the amplitude is dominated by $s$-wave structures and its dependence on the angular variables is weak, so that the zeroth Chebyshev moments carry the bulk of the amplitude.  The MA channel is dominated by $\mathcal{S}^\pm$, the scalar-diquark-like structures, followed by $\mathcal{V}^\pm$; the MS channel is dominated by $\mathcal{A}^+$, the axialvector-diquark-like structure, followed by $\mathcal{V}^\pm$, $\mathcal{P}^\pm$, and $\mathcal{A}^-$.  Hence, the association of MA (MS) with scalar (axialvector) diquark correlations is realised explicitly in the three-body solution.  Within each channel, some components are identically related: $c_{000}(\mathcal{V}^-)=c_{000}(\mathcal{V}^+)$ in the MA channel; and $c_{000}(\mathcal{V}^-)=-c_{000}(\mathcal{V}^+)$, $c_{000}(\mathcal{P}^-)=-c_{000}(\mathcal{P}^+)$ in the MS channel.  Moreover, at vanishing relative momenta, the leading MA and MS components coincide, $c_{000}(\mathcal{S}^+)=c_{000}(\mathcal{A}^+)$, as do $c_{000}(\mathcal{V}^+)$ (MA) and $c_{000}(\mathcal{P}^+)$ (MS), reflecting the permutation symmetry that links the two flavour channels.

All components fall by roughly an order of magnitude between $0.1$ and $1\,$GeV$^2$, and are negligible beyond $p^2,q^2\gtrsim 3\,$GeV$^2$.  Their profiles nevertheless differ in an instructive way.  The leading MA component falls more slowly with the $(12)$-pair relative momentum than the leading MS component -- half maximum at $q^2\approx 0.18\,$GeV$^2$ \emph{cf}.\ $0.09\,$GeV$^2$ -- whereas the ordering is reversed in the dependence on $p^2$, the momentum of the third quark relative to the pair ($0.13\,$GeV$^2$ \emph{cf}.\ $0.18\,$GeV$^2$).  This suggests that the quark pair is more compact in the MA channel than in the MS channel.

\subsection{\texorpdfstring{Quark\,+\,diquark Faddeev equation}{Quark + diquark Faddeev equation}}
\label{S:qdq}
As first revealed, independently, in Refs.\,\cite{Cahill:1988dx, Reinhardt:1989rw, Efimov:1990uz}, a $q(qq)$ picture emerges when one exploits the pairing tendency of fermions to simplify the three-body problem in Fig.\,\ref{FigFaddeev}\,\textsf{A}.  One thereby arrives at a bound-state equation that describes dressed quarks and fully-interacting, nonpointlike quark\,+\,quark (diquark) correlations built therefrom, which bind together into a baryon, at least in part, because of the continual exchange of roles between the spectator and diquark-participant quarks -- Fig.\,\ref{FigFaddeev}\,\textsf{B}.  Many studies of nucleon properties have employed this scheme; see, \emph{e.g}., Refs.\,\cite{Barabanov:2020jvn, Roberts:2020hiw, Ding:2022ows, Cheng:2023kmt, Chen:2023zhh, Yu:2024ovn, Eichmann:2025tzm, Cheng:2025yij}.

Retaining isoscalar-scalar ($[ud]$) and isovector-axialvector ($\{uu\}$, $\{ud\}$, $\{dd\}$) diquark correlations, the nucleon is described by two Poincar\'e-covariant matrix-valued functions \cite{Barabanov:2020jvn, Cheng:2025yij}:
\begin{equation}
\label{qdqamp}
\mathcal{S}(\ell;P) = \sum_{i=1}^{2} s_i(\ell^2,\ell\cdot P)\, \mathcal{S}^i(\ell;P)\,, \qquad
\mathcal{A}_\nu(\ell;P) = \sum_{i=1}^{6} a_i(\ell^2,\ell\cdot P)\, \gamma_5\mathcal{A}^i_\nu(\ell;P)\,,
\end{equation}
which describe the quark\,+\,scalar-diquark and quark\,+\,axialvector-diquark components, respectively.  Here $\ell$ is the quark-diquark relative momentum, $\mathcal{S}^1 = \mathbb{I}_{\rm D}$, $\mathcal{S}^2 = i\gamma\cdot\hat\ell - \hat\ell\cdot\hat P\,\mathbb{I}_{\rm D}$ ($\hat\ell^2=1$, $\hat P^2=-1$), and the six $\mathcal{A}^i_\nu$ are listed, \emph{e.g}., in Ref.\,\cite{Barabanov:2020jvn}.  In contrast to the 128 independent scalar functions required in the three-body treatment, only $8$ are needed here, which greatly simplifies calculations and may deliver additional insights.  The amplitudes satisfy a Faddeev equation whose kernel describes the breakup and reformation of diquark correlations through exchange of a dressed quark.  In Ref.\,\cite{Cheng:2025yij}, QCD-constrained \emph{Ans\"atze} are employed for the dressed-quark and diquark propagators and the diquark correlation amplitudes; this is the principal limitation of the $q(qq)$ approach: its connection with QCD is less direct than that of the three-body framework.  One aim of the comparisons presented herein is to assist in judging the fidelity of the \emph{Ans\"atze} used.

\begin{figure}[t]
\centering
\includegraphics[width=0.48\textwidth]{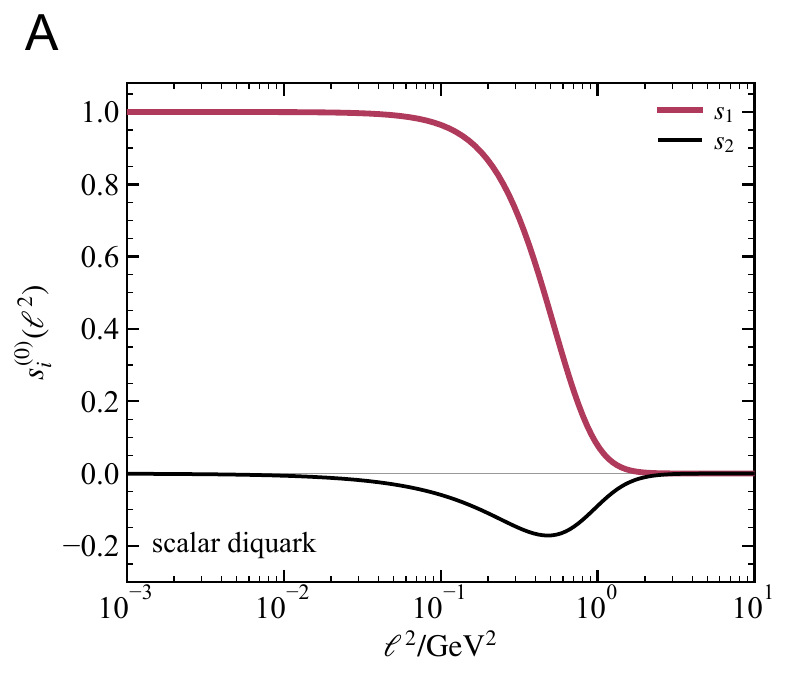}\hfill
\includegraphics[width=0.48\textwidth]{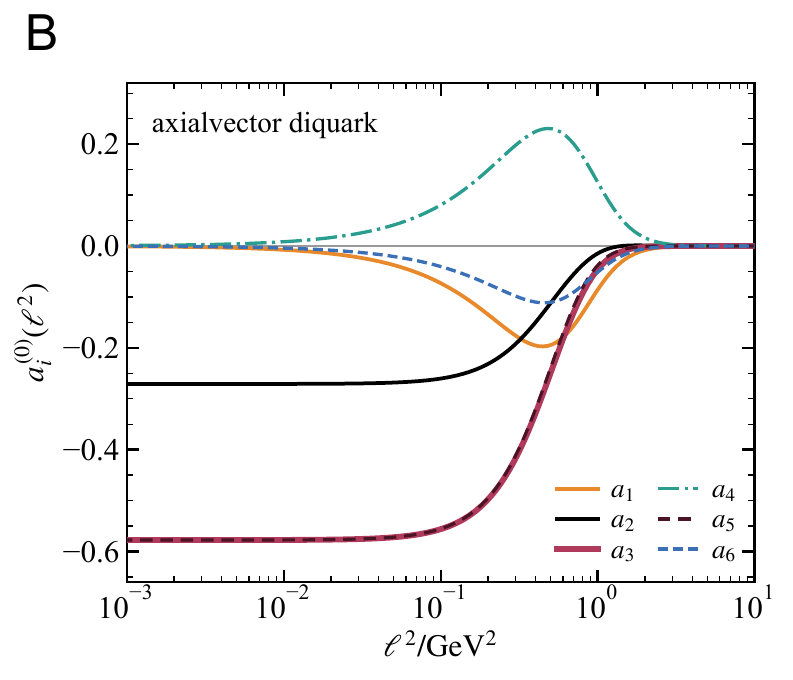}
\caption{Solution of the $q(qq)$ Faddeev equation, Fig.\,\ref{FigFaddeev}\,\textsf{B}: zeroth Chebyshev moments in $\ell\cdot P$, denoted $s_i^{(0)}$, $a_i^{(0)}$, of the functions in Eq.\,\eqref{qdqamp}, as functions of $\ell^2$.  Panel~\textsf{A}: scalar-diquark functions $s_{1,2}$.  Panel~\textsf{B}: axialvector-diquark functions $a_{1,\ldots,6}$, for the $\{uu\}$ correlation in the proton; the $\{ud\}$ functions follow from isospin symmetry, being smaller by a factor $1/\sqrt{2}$ and opposite in sign \cite{Cheng:2025yij}.  $a_3$ and $a_5$ are practically coincident on the domain plotted.  All curves are normalised by $s_1^{(0)}$ at $\ell^2=0$.}
\label{fig:ampqdq}
\end{figure}

The solution is illustrated in Fig.\,\ref{fig:ampqdq}.  As in the three-body case, $s$-wave structures dominate.  Only $s_1$, $a_2$, $a_3$, and $a_5$ are nonzero at $\ell^2=0$; and $a_3=a_5$ there, so that these two terms combine into the structure $\mathcal{A}^3_\nu+\mathcal{A}^5_\nu=\gamma^\perp_\nu$, the component of $\gamma_\nu$ transverse to $P$, which does not depend on the direction of $\ell$.  Relative to $s_1^{(0)}$, one finds $a_3^{(0)}\approx a_5^{(0)}\approx -0.58$ and $a_2^{(0)}\approx -0.27$ at $\ell^2=0$.  The other functions, $s_2$, $a_1$, $a_4$, and $a_6$, multiply structures that do depend on the direction of $\ell$: their zeroth moments vanish as $\ell^2\to 0$ and reach maximum magnitudes of $0.11$--$0.23$ near $\ell^2\approx 0.45\,$GeV$^2$.  The four leading functions share almost the same momentum profile, each falling to half its $\ell^2=0$ value at $\ell^2\approx 0.44$--$0.46\,$GeV$^2$; indeed, $a_3^{(0)}/s_1^{(0)}$ varies by only 2\% on $\ell^2\lesssim 0.5\,$GeV$^2$.  All functions are negligible on $\ell^2\gtrsim 3\,$GeV$^2$.

Owing to the correspondence between the MA (MS) flavour structure of the three-body amplitude and scalar (axialvector) diquark correlations, Figs.\,\ref{fig:amp3b} and \ref{fig:ampqdq} enable a comparison of the two pictures at the level of the bound-state amplitudes.  Both pictures concentrate the amplitude in $s$-wave structures with the Dirac character of scalar and axialvector correlations.  Moreover, since Ref.\,\cite{Cheng:2025yij} uses $\ell=(2p_3-p_1-p_2)/3$, $\ell$ coincides with $p$ in Eq.\,\eqref{Involvesxi}; so the dependence on $\ell^2$ in Fig.\,\ref{fig:ampqdq} can be compared directly with that on $p^2$ in Fig.\,\ref{fig:amp3b}\,\textsf{A}, \textsf{C}.  The $q(qq)$ functions fall more slowly: half maximum at $\ell^2\approx 0.45\,$GeV$^2$ \emph{cf}.\ $p^2\approx 0.13\,$GeV$^2$ (MA) and $0.18\,$GeV$^2$ (MS).  This comparison requires some care.  The $q(qq)$ kernel is built from \emph{Ans\"atze} and yields $m_N=1.18\,$GeV, the mass of the dressed-quark core \cite{Cheng:2025yij}, not $0.94\,$GeV; dependence on the relative momentum within the pair, $q$, is carried by the diquark correlation amplitudes, not by $s_i$, $a_i$; and those correlation amplitudes have different canonical couplings, so the size of $a_i$ relative to $s_i$ does not by itself measure the relative strength of axialvector and scalar correlations.  In the three-body solution, by contrast, the leading MA and MS components are equal at vanishing relative momenta.  We note, too, that the slower $q(qq)$ fall-off is qualitatively in line with the smaller $q(qq)$ Dirac radii in Table~\ref{tab:rad}.  As will become apparent in Sec.\,\ref{S:density}, this correspondence is key to understanding the flavour-separated densities.

\section{Electromagnetic currents and form factors}
\label{S:Current}
For an on-shell nucleon, the electromagnetic current can be written
\begin{equation}
J_\mu^N(Q) = ie\, \Lambda_+(P_f) \Big[ F_1^N(Q^2)\, \gamma_\mu + \frac{1}{2 m_N}\, \sigma_{\mu\nu} Q_\nu\, F_2^N(Q^2) \Big] \Lambda_+(P_i)\,,
\label{NucleonCurrent}
\end{equation}
where $e$ is the positron charge, $P_{i,f}$ are the incoming and outgoing nucleon momenta, $Q=P_f-P_i$, $P_{i,f}^2=-m_N^2$, $\Lambda_+(P_{i,f})$ are positive-energy projection operators, and $F_{1,2}^N$ are the Dirac and Pauli form factors.  The Sachs form factors are $G_E^N = F_1^N - \tau F_2^N$, $G_M^N = F_1^N + F_2^N$, $\tau = Q^2/[4 m_N^2]$ \cite{Sachs:1962zzc}.

\begin{figure}[t]
\centering
\includegraphics[width=0.85\textwidth]{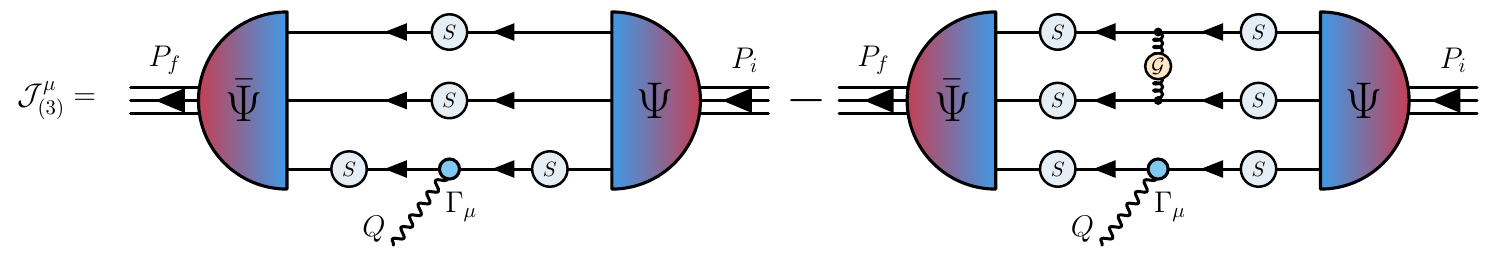}
\caption{The $a=3$ component of the RL nucleon electromagnetic current, Eq.\,\eqref{ImFF}; the $a=1,2$ components follow from symmetries \cite{Eichmann:2011pv}.  $\delta$, $\delta'$ are spinor indices and $\rho$, $\rho'$ are isospin indices; $\Gamma_\mu$ is the dressed photon\,+\,quark vertex.  Adapted from Ref.\,\cite{Xu:2025ups}.}
\label{FigCurrent}
\end{figure}

In the three-body approach, the RL current is obtained from the seven-point Schwinger function depicted in Fig.\,\ref{FigCurrent}.  Following Ref.\,\cite{Eichmann:2011pv}, it is expressed as
\begin{equation}
\label{eq:Jcurrent}
J_\mu^N(Q) = \sum_{a=1}^3\sum_{\rho',\rho=0}^{1} \mathsf{F}^{(a)N}_{\rho'\rho}\, [J^{(a)}_{\rho'\rho}]_\mu\,,
\end{equation}
where the flavour matrices follow from Eq.\,\eqref{FlavourM} and the quark charge matrix ${\rm diag}[2/3,-1/3]$; for $a=3$, $\mathsf{F}^{(3)p} = {\rm diag}[2/3, 0]$ and $\mathsf{F}^{(3)n} = {\rm diag}[-1/3, 1/3]$ in $(\rho',\rho)$ space.  The $a=3$ component of the current is
\begin{align}
[J_{\rho'\rho}^{(3)}]_{\mu}^{\delta'\delta}
= \int_{dp\, dq} & \bar{\psi}^{\rho'\delta'}_{\beta'\alpha'\gamma'}(p_f,q_f,P_f)\,
S_{\alpha'\alpha}(p_1)\, S_{\beta'\beta}(p_2)\,
[\chi_{\mu}(p_3,Q)]_{\gamma'\gamma} \nonumber \\
& \times \big[\psi^{\rho\delta}_{\alpha\beta\gamma}(p_i,q_i,P_i) - \psi^{\rho\delta(3)}_{\alpha\beta\gamma}(p_i,q_i,P_i)\big],
\label{ImFF}
\end{align}
where $p_{f,i} = p \pm Q/3$, $q_f=q_i=q$, $\chi_\mu(p_3,Q) = S(p_3^+)\Gamma_\mu(p_3,Q) S(p_3^-)$, $p_3^\pm = p_3 \pm Q/2$, and the dressed photon\,+\,quark vertex, $\Gamma_\mu$, solves its inhomogeneous RL Bethe-Salpeter equation \cite{Xu:2019ilh, Xu:2021mju}; subtracting $\psi^{(3)}$ avoids double counting the gluon exchange between the spectator quarks.  Using permutation symmetry, one arrives at
\begin{equation}
\label{CurrentNucleon}
J^{p}_{\mu} = 2 [J_{00}^{(3)}]_{\mu}\,, \qquad
J^{n}_{\mu} = [J_{11}^{(3)}-J_{00}^{(3)}]_{\mu}\,.
\end{equation}
Since the calculation is made at the hadron scale, $\zeta_{\cal H}$, whereat all properties of a hadron are carried by its dressed valence degrees of freedom \cite{Yin:2023dbw}, flavour-separated currents may be defined:
\begin{equation}
\label{FlavourSep}
J^{u}_{\mu} = 2 J^{p}_{\mu} + J^{n}_{\mu} = 3 [J_{00}^{(3)}]_\mu + [J_{11}^{(3)}]_\mu \,, \qquad
J^{d}_{\mu} = J^{p}_{\mu} + 2 J^{n}_{\mu} = 2 [J_{11}^{(3)}]_\mu \,.
\end{equation}
Hence, the in-proton $d$-quark form factors receive no contribution from MA components of the proton Faddeev amplitude.  This has a long-known analogue in the $q(qq)$ picture: $d$ quarks cannot be perceived by a hard probe unless axialvector diquark correlations are present in the proton \cite{Close:1988br, Cui:2021gzg, Cheng:2023kmt}.

In the $q(qq)$ approach, the current involves photon couplings to the dressed quark, to the diquark correlations (including scalar\,$\leftrightarrow$\,axialvector transitions), to the exchanged quark, and to the diquark correlation amplitudes (seagull terms) \cite{Barabanov:2020jvn, Cheng:2025yij}.  The form factors of Ref.\,\cite{Cheng:2025yij} are reported as functions of $x=Q^2/m_N^2$; we use $m_N=0.94\,$GeV to express them in GeV$^2$.

In the three-body framework, the canonical normalisation of the Faddeev amplitude is fixed by ensuring $G_E^p(0)=1$; and current conservation is expressed, \emph{inter alia}, in $G_E^n(0)\equiv 0$.  The three-body form factors \cite{Yao:2024uej} have been extended to large $Q^2$ using the Schlessinger point method (SPM) \cite{Schlessinger:1966zz, Schlessinger:1968vsk, Tripolt:2016cya}, which also provides a $1\sigma$ uncertainty band.

\section{Light-front transverse densities}
\label{S:density}
\subsection{Dirac charge densities}
Nucleon light-front transverse charge densities are defined as \cite{Miller:2010nz}
\begin{equation}
\label{Tcharge}
\rho_{ch}^{\mathsf t}(b) = \int\frac{d^2\Delta}{(2\pi)^2}\, {\rm e}^{i \vec{b}\cdot \vec{\Delta}}\, F_1^{\mathsf t}(\Delta^2)
= \int_0^\infty \frac{d\Delta}{2\pi}\, \Delta\, J_0(b\Delta)\, F_1^{\mathsf t}(\Delta^2)\,,
\end{equation}
where $Q^2=\Delta^2$, with $\vec{\Delta}$ purely transverse, $\Delta=|\vec\Delta|$, $b=|\vec{b}|$, $\vec{b} = b\,(\cos\phi_b,\sin\phi_b)$, $J_0$ is a Bessel function of the first kind, and ${\mathsf t}=p,n,u,d$.  Following Eq.\,\eqref{FlavourSep}, $\rho_{ch}^u = 2 \rho_{ch}^p + \rho_{ch}^n$ and $\rho_{ch}^d = \rho_{ch}^p + 2 \rho_{ch}^n$.

\begin{figure}[t]
\centering
\fpanel{A}{$\rho_{ch}^p(b)$}{\includegraphics[width=\linewidth]{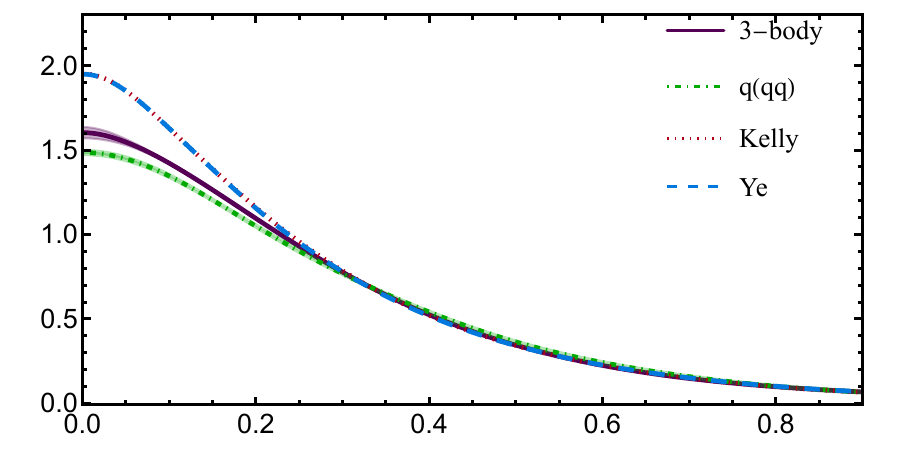}}{$b/$fm}\hfill
\fpanel{B}{$\rho_{ch}^n(b)$}{\includegraphics[width=\linewidth]{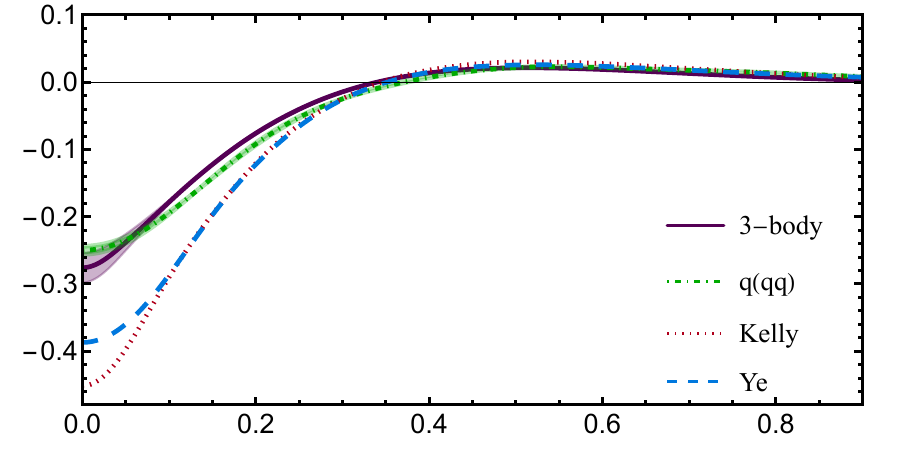}}{$b/$fm}\\[1.5ex]
\fpanel{C}{$\rho_{ch}^u(b)$}{\includegraphics[width=\linewidth]{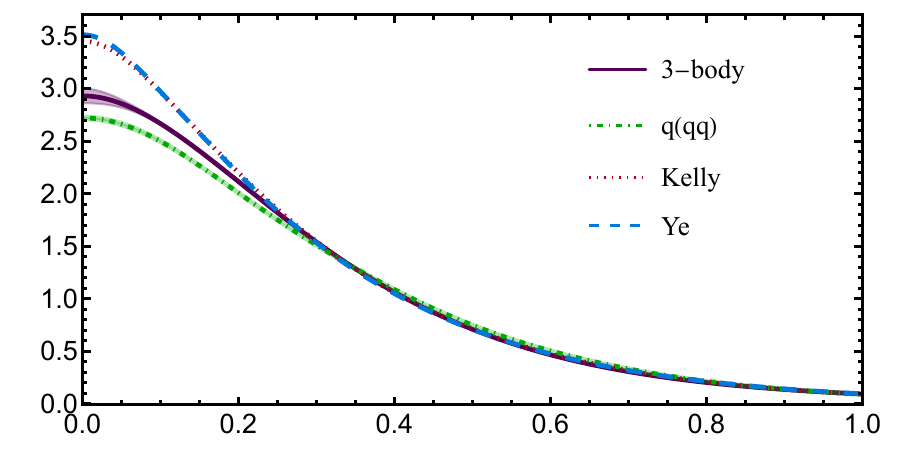}}{$b/$fm}\hfill
\fpanel{D}{$\rho_{ch}^d(b)$}{\includegraphics[width=\linewidth]{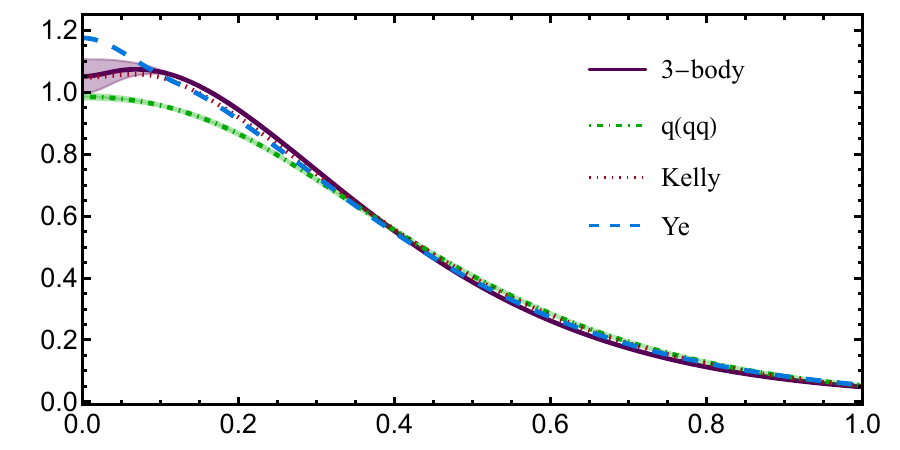}}{$b/$fm}
\caption{Light-front transverse charge densities (in fm$^{-2}$).  Panels~\textsf{A}, \textsf{B}: proton and neutron.  Panels~\textsf{C}, \textsf{D}: in-proton $u$ and $d$ quarks.  Solid purple curve -- three-body Faddeev equation \cite{Yao:2024uej}, with like-colour band indicating the SPM uncertainty; dot-dashed green -- $q(qq)$ \cite{Cheng:2025yij}; dashed blue -- data parametrisation of Ref.\,\cite{Ye:2017gyb}; dotted red -- data parametrisation of Ref.\,\cite{Kelly:2004hm}.  Adapted from Ref.\,\cite{Xu:2025ups}.}
\label{F1pn}
\end{figure}

The densities are displayed in Fig.\,\ref{F1pn}.  In all cases, there is fair agreement between the three-body and $q(qq)$ predictions; and both agree semiquantitatively with results obtained using fits to data as input to Eq.\,\eqref{Tcharge}.  The modest differences between theory and data-based inferences might be connected with the omission of meson-baryon final-state interactions (meson cloud effects) in the calculations; further studies are needed before that can be confirmed.

The proton density is positive definite on the entire $b$ domain; hence, at every point in the transverse plane, the valence $u$-quark density exceeds the valence $d$-quark density.  On the other hand, the neutron density is negative on a sizeable neighbourhood of the neutron's centre of transverse momentum, becoming positive at $b\approx 0.35\,$fm and remaining so thereafter \cite{Miller:2007uy}.  Mathematically, this is a consequence of $F_1^n(0)=0$; the non-positive character of $F_1^n(\Delta^2)$, whose magnitude rises to a maximum and then falls monotonically to zero with increasing $\Delta^2$; and the oscillatory behaviour of the weight $\Delta J_0(b\Delta)$ at fixed $b$.  Thus, on $b\lesssim 0.35\,$fm, the in-proton $u$-quark density exceeds twice the $d$-quark density, a feature that does not persist on the complementary domain.  These expectations are realised in Fig.\,\ref{F1pn}\,\textsf{C}, \textsf{D}; moreover, both in-proton valence-quark densities are positive definite.

\begin{table}[t]
\caption{In-proton valence-quark transverse density radii, defined via Eq.\,\eqref{TransverseRadii} and its analogue for the magnetisation (normalised by $\kappa^{u,d}$), and anomalous magnetic moments, Eq.\,\eqref{Tmagnet}.  The predictions are compared with inferences from data parametrisations \cite{Ye:2017gyb, Kelly:2004hm} and with empirical values \cite{ParticleDataGroup:2024cfk}, flavour-separated using $\kappa_u = 2\kappa_p+\kappa_n$, $\kappa_d = \kappa_p+2\kappa_n$.}
\label{tab:rad}\label{tab:kappa}
\begin{tabular*}{\textwidth}{@{\extracolsep{\fill}}lcccccccc@{}}
\toprule
 & \multicolumn{4}{c}{Radii/fm} & \multicolumn{4}{c}{Anomalous magnetic moments} \\
\cmidrule(lr){2-5}\cmidrule(l){6-9}
 & $\lambda_1^u$ & $\lambda_1^d$ & $\lambda_2^u$ & $\lambda_2^d$ & $\kappa_p$ & $\kappa_n$ & $\kappa_u$ & $\kappa_d$ \\
\midrule
3-body & 0.68 & 0.69 & 0.63 & 0.68 & 1.23 & $-1.33$ & 1.13 & $-1.43$ \\
$q(qq)$ & 0.53 & 0.57 & 0.62 & 0.67 & 1.80 & $-1.86$ & 1.74 & $-1.92$ \\
\midrule
Ye \cite{Ye:2017gyb} & 0.66(1) & 0.67(1) & 0.71(12) & 0.69(14) & & & & \\
Kelly \cite{Kelly:2004hm} & 0.65 & 0.66 & 0.68 & 0.76 & & & & \\
Exp.\ \cite{ParticleDataGroup:2024cfk} & & & & & 1.79 & $-1.91$ & 1.67 & $-2.03$ \\
\bottomrule
\end{tabular*}
\end{table}

These remarks can be quantified via the transverse-plane Dirac radii:
\begin{equation}
\label{TransverseRadii}
(\lambda_1^{\mathsf t})^2 := \frac{\int\! d^2 \vec b\, b^2 \rho_{ch}^{\mathsf t}(b)}{\int\! d^2\vec b\, \rho_{ch}^{\mathsf t}(b)}
= \left. - 4 \frac{d}{dQ^2} \ln F_1^{\mathsf t}(Q^2)\right|_{Q^2=0}\,, \quad {\mathsf t}=u,d\,.
\end{equation}
The predictions are compared with data-based inferences in Table~\ref{tab:rad}.  The three-body radii are noticeably larger than those obtained from the $q(qq)$ form factors, owing to the three-body densities having greater support on $b\gtrsim 1.1\,$fm (equivalently, the $Q^2\simeq 0$ slope of the three-body form factors is greater \cite{Cheng:2025yij}); and they are in better agreement with the data-based inferences.  Overall, the valence $u$- and $d$-quark transverse-plane Dirac radii are practically indistinguishable.

\subsection{Pauli magnetisation densities}
By analogy with Eq.\,\eqref{Tcharge}, light-front transverse magnetisation densities are defined as
\begin{equation}
\rho_{m}^{\mathsf t}(b) = \int_0^\infty \frac{d\Delta}{2\pi}\, \Delta\, J_0(b\Delta)\, F_2^{\mathsf t}(\Delta^2)\,,
\qquad
\kappa^{\mathsf t} = 2\pi \int_0^\infty db\, b\, \rho_m^{\mathsf t}(b)\,,
\label{Tmagnet}
\end{equation}
where $\kappa^{\mathsf t}$ are the anomalous magnetic moments.  Predictions for $\kappa^{\mathsf t}$ are listed in Table~\ref{tab:kappa}.  The three-body values are underestimated; as explained in Ref.\,\cite{Yao:2024uej}, this is a failing of RL truncation, which yields a photon\,+\,quark vertex whose dressed-quark anomalous magnetic moment term is too small.  The weakness is corrected in higher-order truncations \cite{Chang:2010hb}, which have been implemented in studies of mesons \cite{Xu:2022kng} and may, in future, be adapted to baryons.

\begin{figure}[t]
\centering
\fpanel{A}{$\rho_{m}^p(b)/\kappa_p$}{\includegraphics[width=\linewidth]{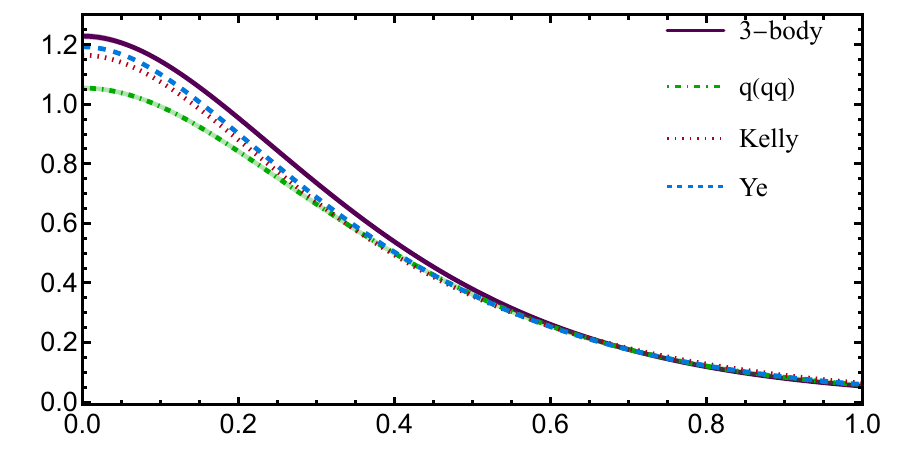}}{$b/$fm}\hfill
\fpanel{B}{$\rho_{m}^n(b)/\kappa_n$}{\includegraphics[width=\linewidth]{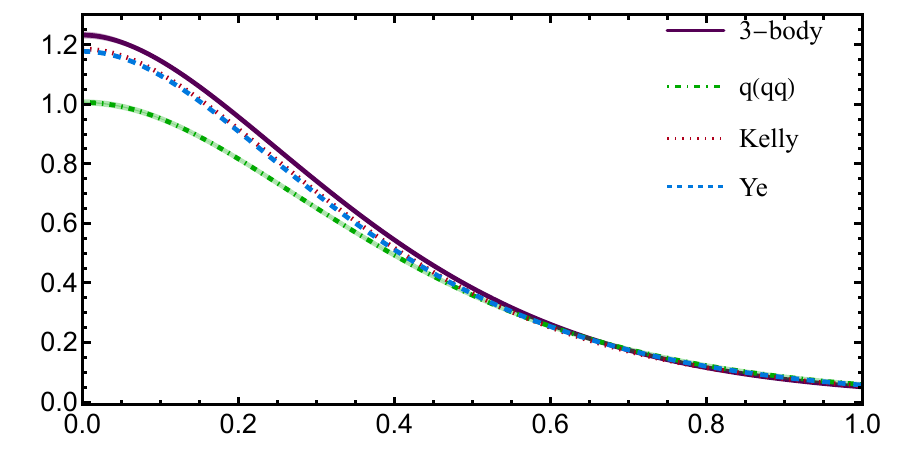}}{$b/$fm}\\[1.5ex]
\fpanel{C}{$\rho_{m}^u(b)/\kappa_u$}{\includegraphics[width=\linewidth]{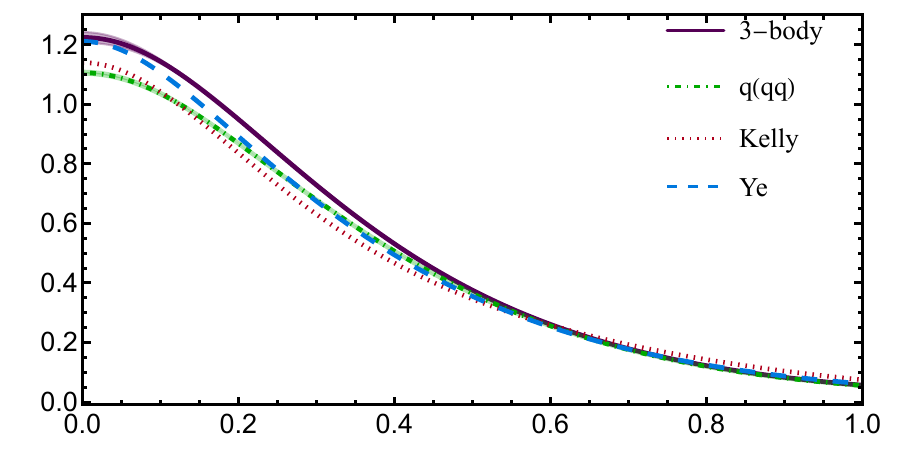}}{$b/$fm}\hfill
\fpanel{D}{$\rho_{m}^d(b)/\kappa_d$}{\includegraphics[width=\linewidth]{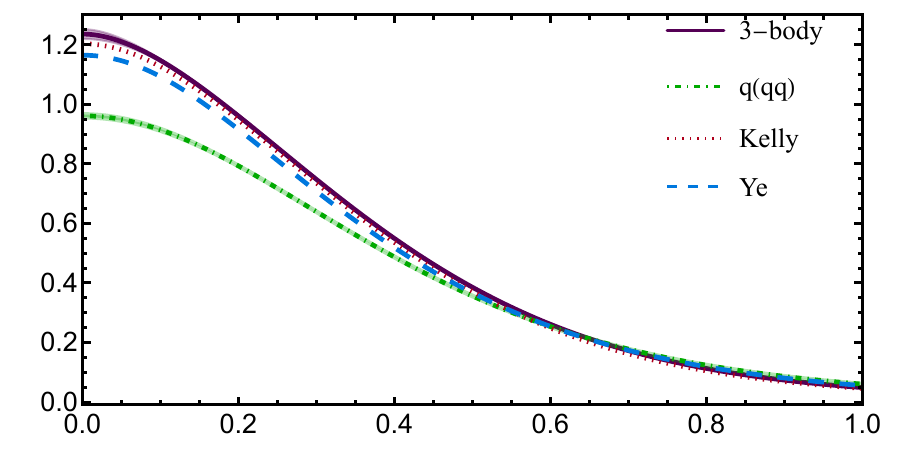}}{$b/$fm}
\caption{Light-front transverse magnetisation densities, Eq.\,\eqref{Tmagnet}, normalised by the anomalous magnetic moments (in fm$^{-2}$).  Panels~\textsf{A}, \textsf{B}: proton and neutron.  Panels~\textsf{C}, \textsf{D}: in-proton $u$ and $d$ quarks.  Legend as in Fig.\,\ref{F1pn}.  Adapted from Ref.\,\cite{Xu:2025ups}.}
\label{fig:F2pn}
\end{figure}

Transverse magnetisation densities are displayed in Fig.\,\ref{fig:F2pn}.  Owing to division by $\kappa$, all are positive definite.  For the proton and neutron, the three-body predictions agree best with the empirical inferences, although the $q(qq)$ results are also a fair match.  Regarding the flavour separation, the valence $d$ quark is magnetically very active within the proton: the unnormalised $d$-quark magnetisation density is approximately equal in magnitude, although opposite in sign, to that of the $u$ quark -- indeed, $|\kappa_d|>|\kappa_u|$, Table~\ref{tab:kappa} -- even though the proton contains two valence $u$ quarks and only one $d$ quark.  This is correlated with indications from the $q(qq)$ picture that, at $\zeta_{\cal H}$, the orbital angular momentum of the valence $u$ quarks in the proton is much smaller than that of the valence $d$ quark \cite{Cheng:2023kmt, Yu:2024qsd, Yu:2024ovn}.  Furthermore, since in a scalar-diquark-only proton the $d$ quark is locked within a magnetically inert $[ud]$ correlation, axialvector diquarks are crucial for understanding $\rho_m^d$.  Translated into the three-body framework, Eq.\,\eqref{FlavourSep}, MS components of the Faddeev amplitude (Fig.\,\ref{fig:amp3b}\,\textsf{C}, \textsf{D}) are essential for a valid description of in-proton valence $d$-quark behaviour.

The transverse-plane Pauli radii, $\lambda_2^{u,d}$, are defined by analogy with Eq.\,\eqref{TransverseRadii}, with normalisation by $\kappa^{u,d}$; they are listed in Table~\ref{tab:rad}.  In this case, the three-body and $q(qq)$ predictions agree, with ordering consistent with the data-based inferences.  Hence, the in-proton transverse-plane valence $d$-quark Pauli radius is $9(2)$\% larger than that of the valence $u$ quark.

\subsection{Transversely polarised nucleon}
For a nucleon polarised transversely to its direction of motion, the light-front transverse charge density receives an additional contribution \cite{Burkardt:2002hr, Carlson:2007xd}:
\begin{align}
\rho_{T}^{\mathsf t}(\vec b) & = \rho_{ch}^{\mathsf t}(b) + \sin(\phi_b - \phi_s)\, \frac{1}{2m_N b}\, \tilde\rho_m^{\mathsf t}(b)\,, \label{SpinChargeDensity1}\\
\tilde\rho_m^{\mathsf t}(b) & = - b\, \frac{\partial \rho_m^{\mathsf t}(b)}{\partial b}
= b \int_0^\infty\frac{d\Delta}{2\pi}\, \Delta^2 J_1(b\Delta)\, F_2^{\mathsf t}(\Delta^2)\,, \label{SpinChargeDensity}
\end{align}
where $\vec{S} = (\cos\phi_s,\sin\phi_s)$ is the polarisation unit vector.  (The relative sign reflects the remarks in Ref.\,\cite[Addendum]{Burkardt:2002hr}.)  With the spin along $+\hat x$ ($\phi_s = 0$), Fig.\,\ref{IMF}, the density is distorted in the $\hat y$ direction: on any domain where $\tilde\rho_m^{\mathsf t}>0$, $\rho_{ch}^{\mathsf t}$ is enhanced on the positive $\hat y$ axis ($\phi_b=\pi/2$) and depleted on the negative $\hat y$ axis.

\begin{figure}[t]
\centering
\fpanel{A}{3-body: $\rho^p(b_x=0,b_y)$}{\includegraphics[width=\linewidth]{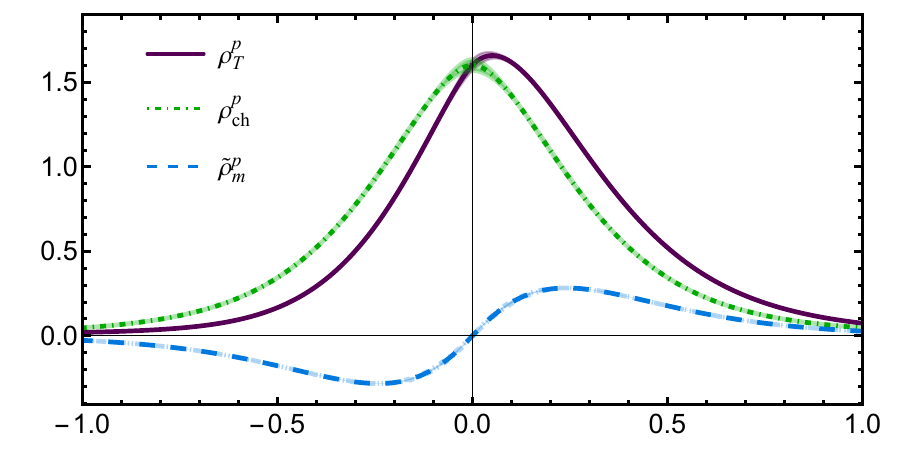}}{$b_y/$fm}\hfill
\fpanel{B}{3-body: $\rho^n(b_x=0,b_y)$}{\includegraphics[width=\linewidth]{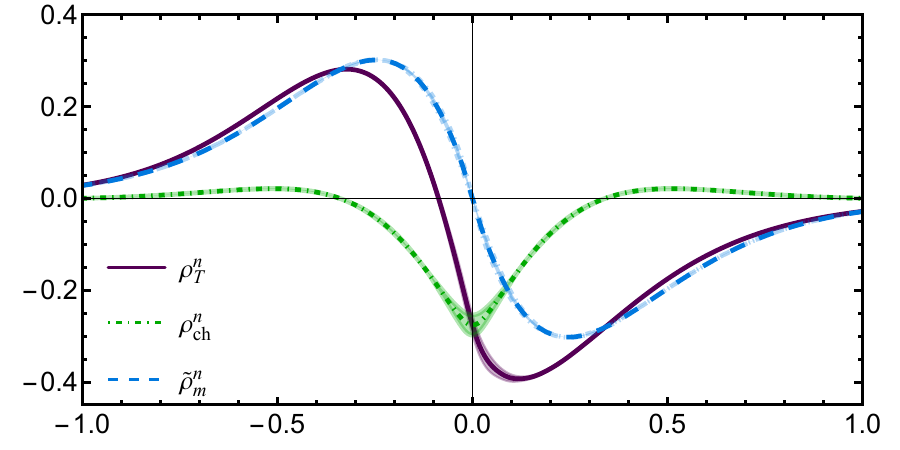}}{$b_y/$fm}\\[1.5ex]
\fpanel{C}{$q(qq)$: $\rho^p(b_x=0,b_y)$}{\includegraphics[width=\linewidth]{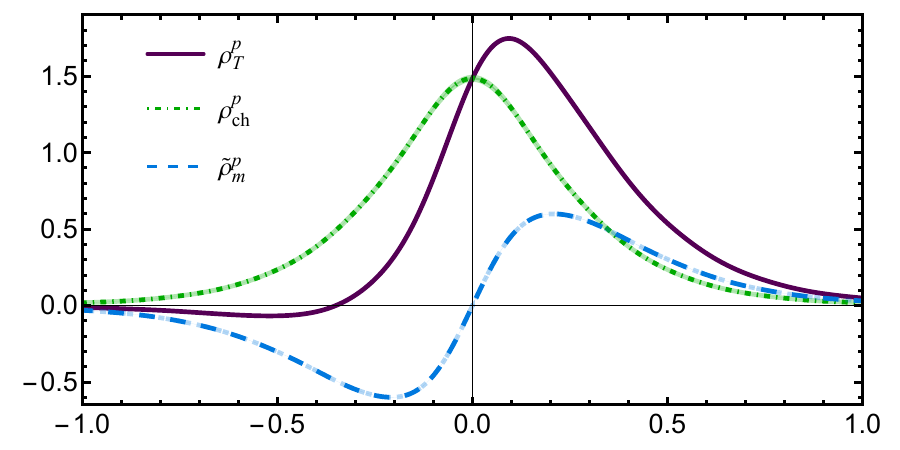}}{$b_y/$fm}\hfill
\fpanel{D}{$q(qq)$: $\rho^n(b_x=0,b_y)$}{\includegraphics[width=\linewidth]{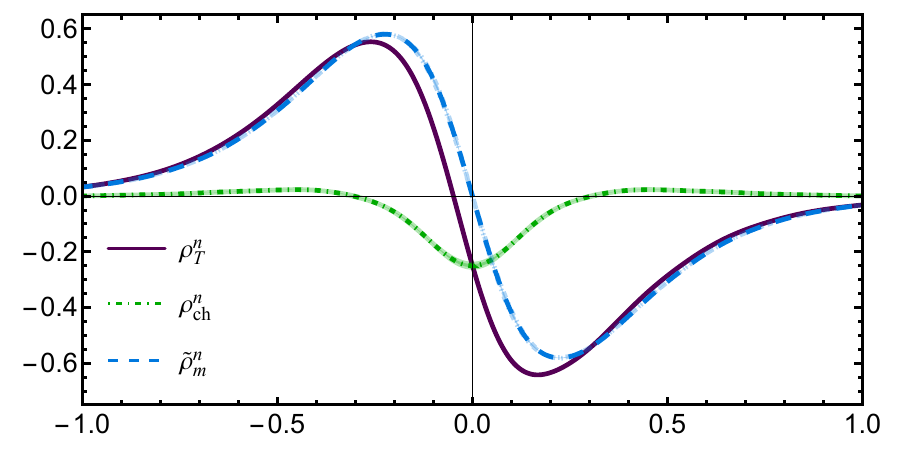}}{$b_y/$fm}
\caption{Proton (\textsf{A}, \textsf{C}) and neutron (\textsf{B}, \textsf{D}) light-front transverse densities for a nucleon polarised along $+\hat x$, evaluated along the $\hat y$ axis.  Panels~\textsf{A}, \textsf{B}: three-body \cite{Yao:2024uej}; Panels~\textsf{C}, \textsf{D}: $q(qq)$ \cite{Cheng:2025yij}.  Solid purple -- $\rho_T$; dot-dashed green -- $\rho_{ch}$; dashed blue -- $\tilde\rho_{m}$.  Like-coloured bands display the (practically negligible) SPM-related $1\sigma$ uncertainties.  Densities in fm$^{-2}$.  Adapted from Ref.\,\cite{Xu:2025ups}.}
\label{fig:F1pnT}
\end{figure}

\begin{figure}[t]
\centering
\fpanel{A}{proton}{\includegraphics[width=0.8\linewidth]{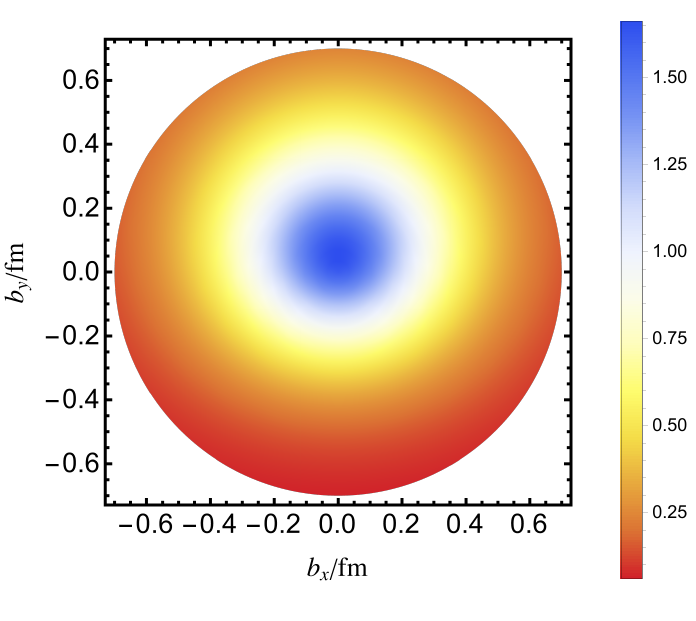}}{}\hfill
\fpanel{B}{neutron}{\includegraphics[width=0.8\linewidth]{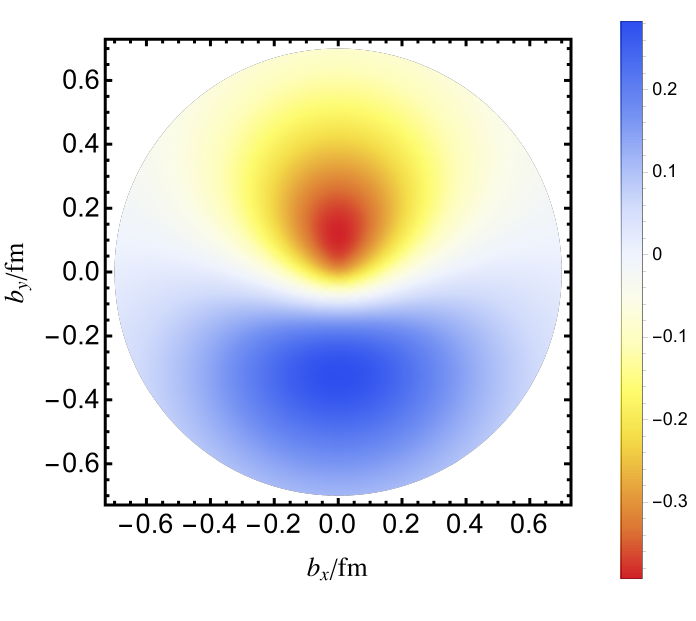}}{}
\caption{Two-dimensional light-front transverse charge densities (in fm$^{-2}$) of the proton (\textsf{A}) and neutron (\textsf{B}) polarised along $+\hat x$, obtained in the three-body framework.  Images for the $q(qq)$ picture are practically equivalent.  Adapted from Ref.\,\cite{Xu:2025ups}.}
\label{fig:F1pnT3d}
\end{figure}

The densities for $+\hat x$ polarised nucleons are depicted in Fig.\,\ref{fig:F1pnT}.  The three-body and $q(qq)$ results are qualitatively and semiquantitatively identical.  There is a depletion of the charge density in the $-\hat y$ direction, with that strength shifted to the $+\hat y$ domain; so, positive charge is shifted toward positive $b_y$ and negative charge toward negative $b_y$.  For the flavour-separated densities \cite{Xu:2025ups}, the $u$-quark density is enhanced on $+\hat y$ and depleted on $-\hat y$; the reverse is true for the $d$ quark, with a larger displacement, which owes to the larger magnitude of $\kappa_d$.  Two-dimensional maps for polarised nucleons are drawn in Fig.\,\ref{fig:F1pnT3d}.  Supposing one works in the infinite momentum frame, these images depict a nucleon moving toward the observer.  Plainly, the distributions are not rotationally symmetric around the direction of motion; instead, positive charge is displaced toward $+\hat y$, whilst reflection symmetry through the $\hat y$ axis is preserved.  Owing to its positive/negative character, the neutron's drift of positive charge into the $+\hat y$ direction is perhaps most noticeable.

\section{Summary and perspective}
\label{S:Summary}
Using input from three-body and $q(qq)$ Faddeev equation results for nucleon elastic electromagnetic form factors, we computed an array of light-front transverse densities for the proton, the neutron, and their dressed valence quarks.  Both frameworks are Poincar\'e-covariant and deliver form factors that extend to very large spacelike $Q^2$, which is crucial for such calculations.  We also compared the bound-state amplitudes in the two pictures, whose MA/MS and scalar/axialvector diquark components are related.  In the three-body amplitude, the MA and MS channels are dominated by $s$-wave structures characteristic of scalar and axialvector diquark correlations, respectively, with the quark pair apparently more compact in the former.  The $q(qq)$ amplitude is likewise dominated by $s$-wave structures, but it falls more slowly with the momentum of the third quark relative to the pair: half maximum at approximately $0.45\,$GeV$^2$ \emph{cf}.\ $0.13$--$0.18\,$GeV$^2$.

The two complementary pictures deliver predictions in qualitative and near-quantitative agreement, which match expectations based on modern parametrisations of data.  In particular: the transverse-plane valence $u$- and $d$-quark Dirac radii are practically indistinguishable, whereas the $d$-quark Pauli radius is roughly 10\% larger; the valence $d$ quark is magnetically far more active than the $u$ quark, a feature that requires MS (axialvector-diquark) components of the nucleon amplitude; and, in a transversely polarised nucleon, the charge densities are no longer rotationally invariant.  Underlying these results is the expression of emergent hadron mass in the quark\,+\,quark interaction that drives both the three-body and $q(qq)$ calculations.

Looking ahead, lattice-QCD results may reach a level of reliability that allows quantitative comparisons with the full array of predictions presented herein.  Of particular relevance to the NSTAR programme, the analysis can be extended to nucleon-to-resonance transition transverse densities.  Such studies are underway in the $q(qq)$ framework; and they will become possible in the three-body approach once the corresponding Poincar\'e-invariant transition form factors have been calculated, an effort which is in train.

\begin{acknowledgements}
We are grateful to C.~D.~Roberts, J.~Rodr\'{\i}guez-Quintero and J.~Segovia for their collaboration on Ref.\,\cite{Xu:2025ups}, and to the organisers of NSTAR 2026.
\end{acknowledgements}

\section*{Statements and Declarations}

\begin{itemize}
\item \textbf{Funding.} Work supported by the Spanish Ministry of Science and Innovation (MICINN) under Grant No.\ PID2022-140440NB-C22.  Z.-Q.Y.\ acknowledges support from Helmholtz-Zentrum Dresden-Rossendorf through the High Potential Programme.  P.C.\ acknowledges support from the Natural Science Foundation of Anhui Province under Grant No.\ 2408085QA028.
\item \textbf{Competing interests.} The authors have no competing interests to declare that are relevant to the content of this article.
\item \textbf{Data availability.} Data will be made available on reasonable request.
\end{itemize}

\bibliographystyle{spphys-arxiv}
\bibliography{NSTAR2026_Xu}

\end{document}